\documentclass[12pt,preprint]{aastex}

\usepackage{epsfig}

\usepackage{xcolor}

\usepackage{amsmath,amssymb,gensymb}

\usepackage{soul}

\begin{document}

\title{The Impact of the Mass Spectrum of Lenses in Quasar Microlensing Studies. Constraints on a Mixed Population of Primordial Black Holes and Stars.}

%\title{(Degeneracy in the) Study of the Abundance and Mass-Function of Primordial Black Holes from Quasar Microlensing}

%\title{Spectrum of Microlens Masses and Magnification Probability Distribution: Degeneracy with the geometric Mean Mass}

%\title{Impact of the {\color{red} Microlenses} Mass-Spectrum on Quasar Microlensing. Application to a Mixed Population of PBHs and Stars.}
%
%
%\title{Quasar Microlensing Induced by a Mixed Population of PBHs and Stars I: Degeneracy with the Geometric Mean Mass
%}

%\title{Mixed Population of PBHs and Stars and Quasar Microlensing: Degeneracy with the Geometric Mean Mass}

%\title{Impact of a Mixed Population of PBHs and Stars in Quasar Microlensing: Degeneracy with the Geometric Mean Mass}
%
%\title{Impact of a Mixture of PBHs and Stars in Quasar Microlensing: Degeneracy with the Geometric Mean Mass}
%
%
%\title{Abundance and Mass-Spectrum of Primordial Black Holes from Quasar Microlensing I: Degeneracy with the Geometric Mean Mass}
%
%\title{Impact of the Mass-Spectrum {\color{red} of a Microlens Population} in Quasar Microlensing Magnification Statistics: Degeneracy with the Geometric Mean Mass}
%
%\title{ Microlensing Magnification Statistics and the Mass-Spectrum of a Microlenses Population: Degeneracy with the Geometrical Mean Mass} 
%
%\title{ Degeneracy with the Geometrical Mean Mass of the Mass-Spectrum of a Microlenses Population} 

%\title{Degeneracy in the Determination from Quasar Gravitational Microlensing of the Microlenses Mass-Spectrum {\color{blue} and Relative Abundances of Stars and Primordial Black Holes}  }

\author{A. ESTEBAN-GUTI\'ERREZ\altaffilmark{1,2}, N. AG\"UES-PASZKOWSKY\altaffilmark{1,2}, E. MEDIAVILLA\altaffilmark{1,2}, J. JIM\'ENEZ-VICENTE\altaffilmark{3,4}, J. A. MU\~NOZ\altaffilmark{5,6}, S. HEYDENREICH\altaffilmark{7}}

%\author{E. MEDIAVILLA\altaffilmark{1,2}, J. JIM\'ENEZ-VICENTE\altaffilmark{3,4}, J. A. MU\~NOZ\altaffilmark{5,6}, H.  VIVES-ARIAS\altaffilmark{5,6}}
%
%
%
\altaffiltext{1}{Instituto de Astrof\'{\i}sica de Canarias, V\'{\i}a L\'actea S/N, La Laguna 38200, Tenerife, Spain}
\altaffiltext{2}{Departamento de Astrof\'{\i}sica, Universidad de la Laguna, La Laguna 38200, Tenerife, Spain}
\altaffiltext{3}{Departamento de F\'{\i}sica Te\'orica y del Cosmos, Universidad de Granada, Campus de Fuentenueva, 18071 Granada, Spain}
\altaffiltext{4}{Instituto Carlos I de F\'{\i}sica Te\'orica y Computacional, Universidad de Granada, 18071 Granada, Spain}
\altaffiltext{5}{Departamento de Astronom\'{\i}a y Astrof\'{\i}sica, Universidad de Valencia, 46100 Burjassot, Valencia, Spain.}
\altaffiltext{6}{Observatorio Astron\'omico, Universidad de Valencia, E-46980 Paterna, Valencia, Spain}        
%%\altaffiltext{7}{Departamento de Estad\'{\i}stica e Investigaci\'on Operativa, Universidad de C\'adiz, Avda Ram\'on Puyol s/n, 11202, Algeciras, C\'adiz, Spain}
%%
%
\altaffiltext{7}{Argelander-Institut f\"ur Astronomie, Auf dem H\"ugel 71, 53121, Bonn, Germany}

\begin{abstract}

We show that quasar microlensing magnification statistics induced by a population of point microlenses distributed according to a mass-spectrum can be very well approximated by that of a single-mass, "monochromatic",  population. When the spatial resolution  (physically defined by the source size) is small as compared with the Einstein radius, the mass of the monochromatic population matches the geometric mean of the mass-spectrum. Otherwise, the best-fit mass can be larger. Taking into account the degeneracy with the geometric mean, the interpretation of quasar microlensing observations under the hypothesis of a mixed population of primordial black holes and stars, makes the existence of a significant population of intermediate mass black holes  ($\sim$ 100$M_\odot$) unlikely but allows, within a two-$\sigma$ confidence interval, the presence of a large population ($\gtrsim 40\%$ of the total mass) of substellar black holes ($\sim$ 0.01$M_\odot$).

\end{abstract}


\section{Introduction \label{intro}}

Quasar microlensing (Chang \& Refsdal 1979; Wambsganss 2006) is very useful to measure the abundance and mass of any population of compact objects of the lens galaxy in a  wide range of masses (see, e.g., Schechter \& Wambsganss 2004, Mediavilla et al. 2009; Pooley et al. 2012; Schechter et al. 2014; Jim\'enez-Vicente et al. 2015a, 2015b, Mediavilla et al. 2017, Schechter 2018, Jim\'enez-Vicente \& Mediavilla, 2019).  {\bf Although most microlensing studies  (see, e.g., Mediavilla et al. 2009, 2017, Schechter et al. 2014 and Schechter 2018) support that the compact objects causing microlensing correspond to the normal stellar population, other works suggest that black holes may also be contributing to the microlens population (Clesse \& Garcia-Bellido 2015, Green 2017, Calcino et al. 2018, Hawkins 2020)}

{\bf In particular}, recent LIGO discoveries related to BH mergers, like the unexpectedly high rates of detection, their unusual masses or their low spins (see, e.g., Kashlinsky et al. 2020) suggest that primordial black holes (PBHs) of intermediate mass ($10M_\odot \lesssim M \lesssim 200M_\odot$) may constitute a substantial part of the dark matter in the Universe and, hence, significantly contribute to microlensing. This hypothesis has been analyzed in Mediavilla et al. (2017) and appears to be inconsistent with current microlensing observations. But the models used in Mediavilla et al. (2017)  are mainly based on a single-mass population of microlenses, which may be inadequate to describe the strongly bimodal distribution of PBHs and stars. 

Could, then, a significant population of PBHs of intermediate mass be hidden in a mix with stars?\footnote{Albeit motivated by LIGO discoveries, this is not the only interesting possibility, we can also consider a bimodal distribution with stars and BHs of substellar mass.}  A direct and comprehensive  study of this possibility multiply the number of cases to be simulated numerically and may always leave open the possibility of including additional parameters or considering more complex cases. Here, we propose an alternative approach, discussing in general the sensitivity of the microlensing magnification statistics (described by the magnification probability density function, PDF($\mu$)) to the existence of a bimodal mass spectrum of compact objects.

The degree of dependence of the microlensing magnification PDF with the spectrum of the microlens masses has been discussed in many works. The initial conjecture that the PDF of a point source microlensed by a population of point masses is independent of the mass spectrum, has been contradicted by several authors (Wyithe \& Turner 2001; Schechter et al. 2004; Congdon et al. 2007). 
{\bf However, Schechter et al. (2004) show that the shape of the mass function of the microlenses is only expected to be important for markedly bimodal} distributions with a large and comparable contribution to the mass density from microlenses of very different masses. Otherwise, for smooth mass functions that span a relatively narrow range of masses, it is usually assumed that the only relevant parameter is the average mass, implicitly taken as the arithmetic mean (AM). However, in contrast with the physical phenomenon under study (microlensing by point masses) the AM is not scale invariant, while the geometric mean (GM) is (see Jim\'enez-Vicente \& Mediavilla 2019).

For a smooth and relatively narrow distribution of masses, the arithmetic  and the geometric means are very similar. However, this is not true for a markedly bimodal distribution of very different masses. Thus, the main objectives of this work are to investigate the role of the GM when the PDF of a bimodal mass spectrum is approximated by the PDF of a single-mass and to study the goodness of this approximation.

These objectives are hampered by a practical question already noticed by Schechter et al. (2014). To approach the point source limit we need to consider magnification maps of very small pixel size with the subsequent reduction of the magnification map size, to maintain an affordable total number of pixels (i.e., of basic computational iterations). This limits the number of high mass microlenses, which, anyway, could not be too large to keep tractable the number of small microlenses.  And this induces a high variability among the PDFs obtained with different random realizations of the positions of the microlenses (sample variance). To avoid this scatter, which would make the comparison between PDFs useless, we use averaged PDFs obtained from a large number (up to 500 in some cases) of random realizations of the spatial distribution of microlenses (i.e., of magnification maps). 

The paper is organized as follows. In \S 2 we introduce the role of the scale-invariant geometric mean mass in the statistical properties of quasar microlensing. In \S 3 we study the magnification statistics of a linear superposition of point masses  discussing the impact of spatial resolution in an idealized scenario (low optical depth). In \S 4 we use numerical simulations to generalize the study to the most realistic case in which cooperative effects among microlenses are present. Finally, in \S 5 we summarize the main conclusions.

%\section{Single-mass analysis of the microlensing statistics of a population of microlenses distributed according to a generic mass-function\label{pdf}}

{
\section{Similarity between magnification maps. Intermediate map between other two. \label{pdf}}
As far as the gravitational potential is scale invariant, we expect gravitational lensing by a single point particle to be also invariant (Schechter et al. 2004). The angle of deflection by a particle of mass $m$,

\begin{equation}
\label{def0}
\vec \alpha ={4Gm\over c^2}{\vec \xi-\vec \xi_0\over (\vec \xi-\vec \xi_0)^2},
\end{equation}
is a power law function of the separation between the light ray and the particle, $\vec \xi-\vec \xi_0$, and hence, it is homogeneous with respect to a length dilation, $\xi \to\xi'= \lambda \xi$. Consequently, we have a mass-length degeneracy (we can conveniently re-scale both, mass and length, leaving invariant the deflection angle), which makes the physics of lensing by a single point particle invariant with respect to mass scaling.

The lens equation for a single point particle of mass $m$, can be written as (see e.g., Schneider, Ehlers \& Falco, 1992),
\begin{equation}
\label{lens0}
\vec \eta{D_d\over D_s}=\vec \xi -{D_{ds}D_d\over D_s}{4Gm\over c^2}{\vec \xi-\vec \xi_0\over (\vec \xi-\vec \xi_0)^2},
\end{equation}
where $\vec \eta$ and $\vec \xi$ are the position vectors at the source and image planes, respectively, and $D_d$, $D_{ds}$ and $D_s$ are the angular diameter distances from the observer to the deflector, from the deflector to the source and from the observer to the source, respectively. Equation \ref{lens0}, and its derivatives, $\vec \nabla_{\vec\xi} \vec \eta$, are invariant under a  transformation of the mass of the microlens, $m \to m' = \lambda m$, if lengths are transformed as $\xi \to \xi' = \sqrt{\lambda} \xi$ ($\eta \to \eta' = \sqrt{\lambda} \eta$). Thus, the magnification map, defined from the scaling factor between surfaces in the lens and image planes, $\mu(\vec \eta)=\sum_I{|\vec \nabla_{\vec\xi} \vec \eta|^{-1}_I}$, where the index $I$ go over all the images of $\vec \eta$, $\vec \xi_I(\vec \eta)$, is also invariant. Consequently, PDF($\mu$), which can be inferred from the magnification map as the fraction of surface that takes the value $\mu$, is as well invariant for a single mass particle. In Appendix \ref{implicit} we formally derive PDF($\mu$) from  $\mu(\vec \eta)$.

When we have a population of microlenses at positions $\xi_i$ with masses $m_i$, Eq. \ref{lens0} becomes,
\begin{equation}
\label{many0}
\vec \eta{D_d\over D_s}=\vec \xi -{D_{ds}D_d\over D_s}\sum_i{4Gm_i\over c^2}{\vec \xi-\vec \xi_i\over (\vec \xi-\vec \xi_i)^2}.
\end{equation}
This equation, and its corresponding magnification map, are also invariant  with respect to a transformation $m_i \to m_i' = \lambda m_i$ if $\xi \to \xi' = \sqrt{\lambda} \xi$ ($\eta \to \eta' = \sqrt{\lambda} \eta$). However, this is not enough to confirm the invariance, as the projected mass density of the microlenses population, $\Sigma$, is also a constraint of the problem. For $n_i$ particles of masses $m_i$, this is given by:
\begin{equation}
\label{sigma0}
\Sigma={1\over \int{d\xi^1d\xi^2} }\sum_i n_im_i,
\end{equation}
which also fulfills the same mass-length degeneracy than Eq. \ref{many0}. That means, that all the microlensing magnification maps corresponding to a fixed mass spectrum, $n_i(m_i)$ that we can build changing the mass of the microlenses, $m_i \to m_i' = \lambda m_i$, but preserving the shape of the mass-spectrum, $n_i(m_i) \to n'_i(m'_i=\lambda m_i)=n_i(m_i)$, are similar (i.e. they are congruent after the length scale transformation, $\xi \to \xi' = \sqrt{\lambda} \xi$).  However, Eq. \ref{many0} and its derivatives, are not invariant with respect to a change in the shape of the mass-spectrum, even if this change leaves $\Sigma$ (Eq. \ref{sigma0}) unaltered.

Let us now consider  two maps of masses $m_1$ and $m_2$. To find a map of mass $m_{GM}$ with intermediate properties between them, we should find a transformation of similarity that in a first step overlaps the map $m_1$ with the map $m_{GM}$ and that applied for a second time overlaps the map $m_{GM}$  with the map $m_2$. That is,

\begin{equation}
 \sqrt{\lambda}=\sqrt{m_1\over m_{GM}}=\sqrt{m_{GM}\over m_2}.
\end{equation}
Thus, $m_{GM}$ is the Geometric Mean (GM), $m_{GM}=\sqrt{m_1m_2}$, which is scale-invariant. If the Arithmetic Mean (AM) of the masses were taken instead, we would have obtained two different scaling factors,

\begin{equation}
\sqrt{\lambda_1}=\sqrt{m_1\over m_{AM}},\ \ 
 \sqrt{\lambda_2} =\sqrt{m_{AM}\over m_2},
\end{equation}
with a ratio betweens scaling factors, 

\begin{equation}
{\sqrt{\lambda_1}\over  \sqrt{\lambda_2}}={m_{GM}\over m_{AM}},
\end{equation}
always less than 1. This means that if we use the non scale-invariant AM we approximate better the properties of the higher mass microlens. In other words, if we are investigating some type of degeneracy of a generic distribution of microlenses characterized by a mass-spectrum with the single-mass case, the single-mass should be reasonably close to the GM  of the mass-spectrum and not to the AM, as it is usually assumed. In many cases this is not important because the GM and the AM are not very different. However, this can be very relevant when we are considering bimodal mass-functions with mass ratios of about 10 {or greater}. {In this context, if the magnification map features corresponding to the smaller mass component are  washed out by progressively lower spatial resolutions (induced by the convolution with the source luminosity profile or by the increase of the pixel size), we expect a transition from the degeneracy with the GM towards the degeneracy with the mass of the larger component. In this case, we should also consider an additional smooth surface mass component to account for the washed out population (Schechter et al. 2014).}

}
\section{Sparse (low optical depth) case.
%: single-mass analysis associated to a bimodal mass-function, extension to a generic mass-function.
}

Better than looking for an intermediate map between other two, we want to study how the statistical properties of a magnification map (i.e. the PDF($\mu$)) in which microlenses of two populations appear mixed, can be mimicked by those of a single-mass map. In principle, the effect of the combination of the two populations is far from linear as cooperative effects between microlenses can appear\footnote{{Even in this general case, it is worth nothing that for particles of masses $m_1 > m_2 > m_3$, the bimodal ($m_1,m_2$) with the same statistical properties than ($m_2,m_3$), is that in which $m_2$ is the GM of $m_1$ and $m_3$.}}. In addition, we should take into account that, in general, the input of the two populations to the total mass may be different. Finally, the spatial resolution can have a different impact in the contribution of the different populations of microlenses to the magnification map depending on their mass. To leave aside the problem of the cooperative effects, we will provisionally suppose that the distribution of microlenses is sparse enough (i.e. that in a region of a given size in Einstein radii around each microlens the deflection of the light rays depends only marginally of the rest of microlenses\footnote{In other words, that above the magnification corresponding to this distance, magnification by a single point lens applies.}).

\subsection{Sparse case in the point source and infinitesimal pixel limit: degeneracy with the single point lens}

We can start considering the limiting case of a point source  (and an infinitesimal pixel). In this limit, if the distribution is sparse enough, the contribution to the average PDF of each microlens will tend to the PDF associated to a single, isolated microlens  which are all similar (because of the mass-length invariance) irrespective of the microlens mass. Consequently, the microlenses mass-spectrum is irrelevant  in this limit and only if the physics of the problem involves a characteristic length, the degeneracy with the single point microlens can be broken. In principle, the most natural choice for this length may be the source size (provided that the source size is greater than the pixel size). However,  in this study we will prefer to focus on the pixel size to try to isolate the statistical properties of the microlenses mass-spectrum from the source properties. In any case, the discretization is nothing but the convolution of the continuous microlensing magnification map, at the points defined by the sampling grid, with a constant source of the size of a pixel. Consequently, the results of discretization can be also interpreted in terms of the presence of a  "pixel source".

%This length can be either the pixel size or the source size (provided that the source size is greater than the pixel size). Although the source size defines a more natural length, in this study we will prefer to focus on the pixel size to try to isolate the statistical properties of the microlenses mass-spectrum from the source properties.
{
\subsection{Sparse case for a finite pixel size: linear superposition of independent point masses and impact of discretization\label{discretization}}

A magnification map is a continuous function of two spatial variables. Its discretization in pixels (or the convolution with the source) implies differences in the PDFs associated to microlenses of different masses breaking the degeneracy of the infinitesimal pixel limit. To illustrate this, we can compare magnification maps obtained for different masses and a fixed pixel size\footnote{Thanks to the mass-length invariance, this is equivalent to compare maps of a same mass but with different pixel sizes}. 

In the low optical depth limit, we can also compare with the linear model for the PDF of the magnification based on the superposition of independent point masses. The general applicability of this model is limited by the presence of caustics generated by the interaction of each microlens with the field created by the rest, but it is quite useful to our purposes. This model has been discussed by many authors (see, e.g., Paczynski 1986, Peacock 1986, Schneider 1987, Kofman et al. 1997) and it is straightforward to adapt it to the context of the magnification maps {\bf (see Appendix \ref{implicit})}. 

In Figure \ref{linear} we have represented, for $\kappa=0.1$, four PDFs corresponding to masses with ratios $1:0.6:0.3:0.1$. To obtain these high S/N PDFs we have averaged histograms corresponding to 500 magnification maps. The maps were built up with a spatial resolution of 0.01 Einstein radii (corresponding to a star of nominal mass equal to 1) per pixel.  In the same Figure we have also represented the linear model {\bf (Equation \ref{final})}, which matches very well (both, quantitative and qualitatively) the global trend even for relatively low magnifications ($\mu\gtrsim 1.6$). The excess bump corresponding to the caustic contributions can be seen around $\mu=2/\kappa$ (Kofman et al. 1997). The effect of the finite spatial resolution is that for high magnifications the PDFs fall below the $\mu^{-3}$ law depicted by the linear approximation model. The departure is more pronounced for the smallest masses. An interesting result, in agreement with our hypothesis, is that the PDF with properties intermediate between the one with the greatest mass (1) and the one with the smallest mass (0.1) is the PDF closest to the geometric mean (0.3). The PDF corresponding to the arithmetic mean (0.6) approximates better the PDF of the largest mass (1).

%Finally, the effect of mass is observed in the tails of the numerically obtained PDFs which fall below the $\mu^{-3}$ law, with a more significant departure for the smallest mass PDF. 

%Obviously, the probability density of an isolated point mass diverges at $\mu=1$. If we, arbitrarily, limit the extent of the continuous magnification map to a radius $y_{max}$, i.e, $\mu_{min}={y_{max}^2+2 \over y_{max}\sqrt{y_{max}^2+4}}$, we obtain the plot of Figure ??? where we have taken $y_{max}=1$. In this Figure we have also represented the PDFs obtained from two magnification maps of pixel sizes of 0.02 and 0.01 Einstein radii. The effect of pixel discretization is to lower the probabilities of high magnifications. 
In summary, for a fixed pixel size, the high magnification tail of the  PDF associated to a small mass particle will depart in a greater extent from the linear solution than the one associated to a bigger mass. As we will see later, the numerical calculations also demonstrate that this effect of lowering the probabilities of high magnifications when decreasing the mass of the microlenses\footnote{Or the spatial resolution for a constant mass by virtue of the mass-length invariance.} is also present for large optical depth.
}
\subsection{Single-mass analysis associated to a bimodal mass-function. Extension to a generic mass-function.}

Once the degeneracy with the point size lens is broken by the discretization (i.e., by the presence of a "pixel size" source), the contribution (i.e. the surface coverage) of each microlens to the map (and hence to the PDF) will be, according to the mass-length invariance, proportional to its mass. Thus, if we have two different populations with masses $m_1$ and $m_2$ and fraction of mass in microlenses $\kappa_1=n_1m_1$ and $\kappa_2=n_2m_2$, the relative contributions of each population to the map are, $m_1^{\kappa_1/(\kappa_1+\kappa_2)}$ and $m_2^{\kappa_2/(\kappa_1+\kappa_2)}$, respectively. Consequently, the compromise mass representing the bimodal distribution will be,

\begin{equation}
m_{GM}=m_1^{\kappa_1/(\kappa_1+\kappa_2)} m_2^{\kappa_2/(\kappa_1+\kappa_2)}.
\end{equation}
 This expression can be extended to include the case of a generic mass-function,

\begin{equation}
m_{GM}=\prod_i m_i^{\left({\kappa_i\over \sum_i \kappa_i}\right)}.
\end{equation}

%{\bf \subsubsection{Sparse case for finite source size}}

%\subsubsection{Impact of the source size}

Finally, it is important to comment that the impact of the discretization (interpretable as the presence of a "pixel source") in a magnification map corresponding to a bimodal mass-function can be very important. In fact, when the pixel size is much smaller than the Einstein radius associated to the mass of one of the populations but larger than the Einstein radius corresponding to the other population, the prints of this last population in the magnification map can be smeared out. 
%
%Following Schechter, Wambsganss \& Lewis (2004) it would be convenient, in this case, to introduce an additional smooth component to account for the mass of the washed out population.

\section{Non-sparse case. Numerical Simulations.}

When we are not limited to the low mass density case, the cooperative effects among microlenses make the previous analysis idealized. A study based on numerical simulations will help us testing to what extent the PDF corresponding to a generic mass-function can be approximated by that of a single-mass function and to study if the GM hypothesis for the mass of the "monochromatic" distribution holds. 
%{\color{blue} , to understand how the single-mass departs from the GM when the source size is comparable to the Einstein radii of part of the masses, and to ascertain what impact has this in the splitting of the total mass between the microlenses and the smooth matter component.} 

According to previous studies, the assumption that the microlenses mass distribution is degenerated with the single-mass case holds for any smooth mass function and there is agreement (Wyithe \& Turner, 2001; Schechter, Wambsganss \& Lewis 2004; Congdon et al. 2007) in that only markedly bimodal distributions with a large and comparable contribution to the mass density from microlenses of very different masses may break the degeneracy  and, even so, only for images that maximize the effects of this bimodality (saddle-point images of high magnification).   Thus, we are going to simulate the PDF of a bimodal mass-spectrum (mock data) and look for the PDF corresponding to a single-mass function that better fits the mock PDF. We start considering{\bf , for convenience,} the case\footnote{\bf This case is unrealistic but useful to to exemplify the impact of spatial resolution.} $\kappa=0.55$ and $\gamma=0$ to study in this computationally less expensive case the impact of pixel size (i.e. of the spatial resolution), and then we consider the extreme case of a saddle-point image of high magnification ($\kappa=\gamma=0.55$).

%
%\subsubsection{\bf \st{Preliminaries: approximating the "point-source" case with a magnification map of finite pixels}}
%
%{\bf \st{What pixel size is needed to guaranty that the magnification histogram is a good approximation,  just to a given magnification, to the theoretical one corresponding to a point-like source?}}

\subsection{Mock PDFs representing the bimodal mass-function}

The relevant parameter to study the bimodal distribution is the ratio between masses, $m_1/m_2$. We compute magnification maps (see Appendix \ref{maps}) for bimodal mass distributions with mass ratios: $m_1/m_2=\{6.25,12.5,25,50,100\}$. Then, we can scale the masses using the mass-length invariance. For convenience, we will assume a fiducial mass $m_2=0.01$ for the small mass stars and, consequently, $m_1=\{0.0625,0.125,0.25,0.5,1\}$. Nevertheless, to offer a physically interesting example, we can adopt $m_2= 0.3 M_\odot$. Then, for a typical lens ($z_l\simeq0.5$, $z_s\simeq2.0$) we have an Einstein radius $\eta(0.3M_\odot)\simeq 10\,\rm light-days$. We start considering that both components have the same contribution to the total mass density, $\kappa_1=\kappa_2=0.55/2$ and $\gamma=0$. In a second step (see \S \ref{gamma}) we consider the $\gamma=0.55$ case. 

%We obtain 200 magnification maps for each one of these 5 mass ratios. Averaging each collection of 200 magnification maps histograms, we obtain 5 PDFs, which are our mock data. 

\subsection{Model PDFs of the single-mass function}

To compare with the above mock data, we compute another set of magnification maps (models) based on single-mass populations to determine the likelihood of these models of reproducing the statistical properties corresponding to the bimodal mass-function. 
%As the previous theoretical study for the sparse case suggest, when the impact of the smoothing with the source is low, the mass of the microlenses should approximate the GM, $m_{GM}=\sqrt{m_1m_2}$. Consequently we will 
We adopt, for the single-mass component, fiducial masses from $m=0.01$ to $m=0.2$ in linear steps of 0.01 (i.e. masses from 0.3 to 6$M_\odot$ in linear steps of 0.3$M_\odot$, in the case of the physical example above considered, $m_2= 0.3 M_\odot$). 
%{\color{blue} In the bimodal case, when the effect of the smoothing is strong respect to the small mass but negligible respect to the large mass, we expect a great part of the contribution to the PDF from the small microlenses to be washed-out transforming its effect in that of a smooth component. For this reason, we are going to consider that the contribution to the total mass of the single-mass microlenses can range between 0.05 and 0.55 according to: $\kappa=\{0.05,0.1,0.15,0.2,0.25,0.3,0.35,0.4,0.45,0.5,0.55\}$. Thus, we have $5\times 11$ maps that after convolution with the corresponding value of $r_s$ will be compared with the ones corresponding to the bimodal mass-function. }

Using the Inverse Polygon Map technique (Mediavillla et al. 2006, 2011),  we  generate (see Appendix \ref{maps}) 200 different magnification maps corresponding to different random positions of the microlenses, for each set of mock ($m_1/m_2$, $\kappa$, $\gamma$) or model ($m$, $\kappa$, $\gamma$) parameters. Finally, the 200 histograms of the magnification maps are averaged in order to calculate a mean histogram and its standard deviation for each mock ($\hat h_i(m_1/m_2)$, $\hat \sigma_i(m_1/m_2)$) and model ($h_i(m)$, $\sigma_i(m)$). The index $i$ runs through all the histogram bins. To avoid edge effects, 200 border pixels of the magnification maps are removed.

\subsection{Comparison between PDFs, best fit and results ($\kappa=0.55,\gamma=0$)}

To assess the similarity between the PDFs, we can use a $\chi^2$  test. For each histogram of the mock data, $\hat h_i(m_1/m_2)$ we can compute $\chi^2(m;m_1/m_2)$ as,

\begin{equation}
\label{chi2s}
\chi^2(m;m_1/m_2)= \sum_i{{\left(\hat h_i(m_1/m_2)- h_i(m)\right)^2\over \hat \sigma^2_i(m_1/m_2)+ \sigma^2_i(m)}},
\end{equation}
where $h_i(m)$ is the histogram of the single-mass model and the index $i$ corresponds to the different bins of the histogram. 
The dependence of  $\chi^2(m;m_1/m_2)$ with mass, $m$, is shown in Figure \ref{all}. In the same Figure we show the best fits for each one of the 5 mock data histograms and the likelihoods of the models,

\begin{equation}
p(m;m_1/m_2)\propto e^{-{1\over 2}\chi^2(m;m_1/m_2)}.
\end{equation}
The errors in the approximation of mock PDFs by best-fit model PDFs are small. The best-fit single-mass distribution approximates the bimodal with average errors ranging from 0.04\% to 0.13\% when the mass ratio $m_1/m_2$ of the bimodal ranges from 6.25 to 100, and the worst-case error from 6\% to 18\% in the same interval of mass ratios.

In Figure \ref{mock_error} we compare the best-fit mass with the geometric mean. We have considered 3 different pixel sizes to take into account the effect of the finite pixel size discussed above. The considered pixel sizes correspond to {\bf 0.0011, 0.0022 and 0.0044 Einstein radii (i.e. to 0.125, 0.25 and 0.50 light-day for a $30M_\odot$ star}).

The results clearly support the degeneracy with the GM modulated by the map resolution (or, physically, by the source size). For the highest resolution the best-fit masses closely match the geometric mean. However, they progressively take values larger than the GM when the resolution is lowered by a factor of two or four. 

%Thus, we have three regimes: high resolution (or very small source) in which best fit masses will fall below the geometric mass (likely because the point source limit is approached), intermediate resolution (or source very small as compared with the Einstein radius) in which the best fit mass will be close to the GM  one and low resolution (or source size comparable to a fraction of the Einstein radius) for which the best fit mass will be greater than the GM.

Notice that, as discussed in \S \ref{discretization}, the impact of a given pixel (or source) size is different for different masses. For this reason if we look in Figure \ref{mock_error}  to the points corresponding to the resolution that better matches the GM, we see that while the lower mass ratios are clearly above the GM, the higher ones match better the GM (a similar trend with the separation from the GM can be seen for the points corresponding to the other two resolutions). This is a consequence, once more, of the mass-length degeneracy: a given pixel size corresponds to a smaller fraction of the Einstein radius for larger masses and, hence, the impact of pixel size (moving the points above the GM curve) is less noticeable for the larger masses. In any case, it is clear from Figure \ref{mock_error} that, leaving aside the modulation induced by the pixel (or source) size, the mass-spectrum is close to the GM and far away from the AM.

These results are based on a particular definition for the "distance" between the histograms given by Eq. \ref{chi2s}. However, there are many other different possible definitions of distance between two histograms. It is interesting to define a family of distances,  $D^2_\alpha$, to check the dependence of the similarity between histograms with the choice of measure. With this aim, we compute $D^2_\alpha(m;m_1/m_2)$ as,

\begin{equation}
\label{measure}
D^2_\alpha(m;m_1/m_2)= \sum_i{{\left(\hat h_i(m_1/m_2)- h_i(m)\right)^2\over \left({\hat h_i(m_1/m_2)+h_i(m)\over 2}\right)^\alpha}}.
\end{equation}
Notice that the parameter $\alpha$ can be used to regulate the relevance of the different bins of PDF($\mu$) in the measure of distance.  
We obtain quite similar values for $\alpha=1$, which is alike to Pearson's $\chi^2$ test and for $\alpha=2$, which gives more weight to the high magnification bins. However, the uncertainty in the determinations of the maximum likelihood estimates is greater for $\alpha=1$. In fact, in the case of the Euclidean distance, $\alpha=0$, the results are rather erratic. This is not strange, as the Euclidean distance strongly enhances the contribution of the low magnification bins, which, as we know from \S \ref{discretization} are quite insensitive to the microlens mass. In fact, if we restrict in Eq. \ref{measure} the sum to bins with microlensing magnification larger than 1 (i.e., $\Delta m_{micro}<0$, in magnitudes), the results obtained with the Euclidean distance also match those obtained with Eq. \ref{chi2s}. This is an interesting result by itself: regions of microlensing magnification larger than 1 are much more sensitive to the effects of changing the microlens masses than regions with no magnification or demagnification. And this result is not  unexpected if we take into account that, by virtue of the length-mass degeneracy, an increase in mass is alike with an increase in spatial resolution, which, obviously affects more to the high magnification regions, which are associated to larger gradients in the magnification maps. This also confirms the results of the low optical depth case. At any rate, in our experience, a correct determination of the histograms in the low magnification bins, needs a huge computational effort (see Appendix \ref{maps}) and any result related to this region of the histograms should be regarded with special caution.

To explore the impact of the high magnification tail in the previous results, we have also limited, in the previous equations, the sum to bins with magnifications in the $-2<\Delta m_{micro}<0$ interval, obtaining results very similar to those obtained in the whole interval, although with larger uncertainties.

\subsection{The extreme bimodal case ($\kappa=\gamma=0.55$)\label{gamma} }

Let us now consider a limiting case, a saddle point image of high magnification  ($\kappa=\gamma=0.55$), which strongly enhances the effects of bimodality (Schechter et al. 2004). In Figure \ref{mock_gamma} we compare the GM, the AM and the best-fit results for this high magnification case. We have also adopted the same 5 mass ratios $m_1/m_2=\{6.25,12.5,25,50,100\}$ considered above. Details about the calculations can be found in Appendix \ref{maps}. It is clear, also in this extreme case, the good approximation provided by the single mass model with a mass corresponding to the GM. 

\section{Discussion: constraints on a mixed population of stars and PBHs}

{\bf The main results of the previous sections are that the PDF($\mu$) corresponding to a single-mass microlens distribution can approximate rather well the PDF($\mu$) of a bimodal distribution, and that the best-fit mass of the single-mass distribution matches the GM of the bimodal (provided that the spatial resolution is small as compared with the Einstein radius of the smallest microlenses). These results, which simplify the study of a mix of stars and BHs, are robust with respect to the criteria used to compare the PDFs. We have even considered an extreme bimodal case with mass ratio of 1/100, and for one type of image that maximize the effects of this bimodality (a saddle-point image with  $\kappa=\gamma=0.55$, Schechter et al.  2004), showing that these conclusions are applicable to virtually any sensible smooth mass function. Ideally, more extreme mass ratios ($<$1/100) and other values in the ($\kappa$,$\gamma$) plane may be explored, but it seems that the selected parameters are representative of the mix of stars and intermediate mass BH.}

{\bf Thus,} if we apply the results of the previous sections to the single mass study of a hypothetical mix of PBHs with normal stars, we would expect that the best-fit mass would be, at least, the geometric mean. In fact, as far as quasar source sizes are supposed to be comparable or larger than  $\sim1\,\rm light-day$\footnote{\bf Notice that, at visible wavelengths, the quasar size can be of several light-days (see, e.g., Fian et al. 2016).}, i.e., greater than the pixel size adopted here  (for a reasonable value of the stellar mass, $m_2=0.3 M_\odot$), we can expect that the monochromatic mass estimates based on observations are even greater than the GM. 

We can use this result to explore the likelihood of a mixed population of PBHs and stars according to microlensing observations {\bf available in the literature. Specifically, we can reinterpret the data and results from Mediavilla et al. 2017 but considering now that,  according to the present work, the masses inferred by these authors using a single-mass function should be taken as a lower limit of the GM of the bimodal mix of stars and BHs. Thus, } if we consider a bimodal distribution with stars of 0.1$M_\odot$ (conservative small value to lower the GM) and 100$M_\odot$ PBHs, the monochromatic best-mass fit should be greater than 3$M_\odot$. If we locate this mass in the likelihood function, $L(\alpha,M)$, for the mass fraction, $\alpha$, and (best-fit) mass, $M$, of the monochromatic population of microlenses, inferred by Mediavilla et al. (2017) from the optical microlensing observations (Figure 1 of these authors), we find a very low probability. Thus, the {\bf reinterpretation} of {\bf the} microlensing observations {\bf and results presented in} Mediavilla et al. (2017)  does not support the presence of a large population of massive PBHs mixed with the normal stars, even when a mass-spectrum is considered for the distribution of the microlenses. 

Very interestingly, the case for substellar mass PBHs is not discarded, within a $2\sigma$ confidence  interval, by microlensing observations. If we consider now a bimodal distribution with PBHs of 0.01$M_\odot$ and stars of 0.3$M_\odot$ (a conservative large value to increase the GM), the resulting GM is $\sim 0.05 M_\odot$, which falls within the $2\sigma$ confidence level in the likelihood function inferred from optical observations by Mediavilla et al. (2017, see Figure 1 of these authors), and corresponds to a range of fraction of mass in microlenses from 20 to 40\%. This GM will also fall within the $1\sigma$ confidence interval of the likelihood function corresponding to X-ray observations (see Figure 2 of Mediavilla et al., 2017), although in this case the fraction of mass in microlenses will be less than 40\%.

Notice that the GM is a lower limit and that the increase of the best-fit mass of the monochromatic population above the GM due to the convolution with the quasar source, may change the likelihood. In the case of the small mass PBHs an increase of the likelihood can be expected. On the other hand, the convolution with the source brightness profile may wash-out in the magnification maps the features of microlensing induced by the small mass component, thus making its contribution to the total magnification similar to that of a smooth matter distribution (see, e.g., Schechter et al. 2014).  Consequently, a thorough analysis of the likelihood of a mixed population of stars and PBHs needs to include, in addition to the single-mass population, a smooth matter component which can account for the washing-out of the small mass population due to the finite size of the source. This approach is computationally very time consuming and we will defer it for a forthcoming work.

\section{Conclusions}

{\bf In the context of the recent detection by LIGO of a population of intermediate ($\lesssim 100\rm\, M_\odot$) mass BHs, we study} the impact of the mass-spectrum of the lenses in the magnification statistics of quasar microlensing {\bf considering} a mixed population of stars and PBHs. {\bf To avoid particularizing to a specific model, we discuss in general the sensitivity of microlensing magnification statistics to a bimodal distribution of compact objects.} We {\bf study} both, the low optical depth case in which a theoretical approach is possible and the non sparse case for which numerical simulations are needed. We take special care in the calculations to avoid the noise induced by edge effects on the magnification maps and by sample variance.  The main conclusions are the following,

1 - In the low optical depth case, {\bf we verify that the PDF with properties intermediate between those of the PDFs related to two different masses, corresponds to the geometric mean of the two masses. \bf It is also worth to mention that} the average of 500 different realizations of the numerical PDFs in order to avoid sample variance, highlights with unprecedented S/N ratio, the excellent matching of the linear model and the departures induced by the spatial resolution in the high magnification tail.

2 - {\bf Numerical simulations, needed to consider the high optical depth case, show that} a single-mass distribution can approximate very well the statistical properties of a strongly bi-modal case\footnote{And, presumably, of any smoother mass function.} with a mean (worst-case) error $\lesssim  0.1$\% ($\lesssim 10\%$) in the range of studied mass ratios (from 6.25 to 100, with convergence $\kappa=0.55$). {\bf This result strongly simplifies the study of a mixed population of stars and BHs.}

3 - The simulations also support that the mass of the "monochromatic" case that best fits the bi-modal distribution is close to the geometric mean of the bi-modal distribution\footnote{\bf This conclusion is robust with respect to the choice of distance between histograms although the low magnification bins ($\Delta m_{micro} \gtrsim 0$) are less sensitive to the change of the mass of the "monochromatic" model. In any case, the results are not critically dependent on the high magnification tail of the distribution ($\Delta m_{micro} \lesssim -2$).}, as suggested by the mass-length invariance. This result depends on the spatial resolution (limited in our case by the pixel size). If the pixel size is small compared with the Einstein radius, the best-fit mass will match the GM. However, if the pixel size is large enough the best-fit mass will take values above the GM.  We interpret this as a selective washing out (induced by the loss of spatial resolution) of the imprints of the small mass component of the bi-modal distribution (Schechter et al. 2014). 

4 - As far as quasar source sizes are supposed to be comparable or larger than  $1\,\rm light-day$, i.e., larger than the pixel sizes adopted here (for a reasonable value of the stellar mass, $m_2=0.3 M_\odot$), we can expect that the mass estimates based on observations are at least the geometric mass or even greater. Consequently,  the results derived from microlensing observations by Mediavilla et al. (2017)  {\bf can be reinterpreted taking the geometric mass as a lower limit of the mass of the "monochromatic" distribution. We find that microlensing observations} do not support the presence of a large population of massive PBHs even when a mass-spectrum is considered for the distribution of the microlenses. However, a significant contribution from substellar BHs can not be discarded within a $2\sigma$ confidence  interval, although simulations including a smooth mass component are needed to have a clearer picture.

\acknowledgements{{\bf We thank the referee for useful comments that significantly improved the quality of the paper.} This research was supported by the Spanish MINECO with the grants AYA2016-79104-C3-1-P and AYA2016-79104-C3-2-P.  J.J.V. is supported by the project AYA2017-84897-P financed by the Spanish Ministerio de Econom\'\i a y Competividad and by the Fondo Europeo de Desarrollo Regional (FEDER), and by project FQM-108 financed by Junta de Andaluc\'\i a. AEG thanks the support from grant FPI-SO from the Spanish Ministry of Economy and
Competitiveness (MINECO) (research project SEV-2015-0548-17-4 and predoctoral
contract BES-2017-082319).}

\appendix
{
\section{Explicit solution of PDF($\mu$) as a function of  $\mu(\vec \eta)$ \label{implicit}}

The probability density function of the microlensing magnification, PDF($\mu$), is proportional to the surface of the magnification map that takes this value, ${\rm PDF}(\mu)d\mu\propto dS(\mu)$.  Let  us  now consider  a  collection  of  open  sets $V_i$  that cover the source plane excepting the caustics, $\cup V_i=R^2-\{C\}$, and are pairwise disjoints, $V_i \cap V_j=\emptyset,\, \forall i, j$. If  it  is  possible  to  do  this source-plane  {\it partition} which leaves  the  caustics outside  the  open  sets,  the magnification map, $\mu(\vec \eta)$, is a continuously differentiable function in each one of the open sets,
\begin{equation}
\mu: V_k \to R, \, \vec \eta \to \mu(\vec \eta),
\end{equation}
and we can apply the implicit function theorem. According to this theorem,  for each open set of the partition, it is possible to obtain an explicit solution  $\vec \eta_i(l,\mu_0)$, of the curve defined by $\mu(\vec \eta)-\mu_0=0$ (i.e., the contour line $\mu=\mu_0$), where $l$ is a parameter, which we will identify with the length of the curve. Thus, the surface of the open set, $V_i$, where the magnification map takes the value $\mu_0$ is, 
\begin{equation}
\label{set}
dS_i(\mu_0)=\int_{\vec \eta_i(l,\mu_0)}{|d\vec \eta_\bot|}\, dl,
\end{equation}
where $dl=\sqrt{(d\eta^1)^2+(d\eta^2)^2}$ and $d\vec \eta_\bot$ is an infinitesimal displacement perpendicular to $\vec \eta(l,\mu_0)$. As $\mu$ is differentiable, we write, 

\begin{equation}
d\mu=\vec \nabla \mu \cdot d\vec\eta=|\vec \nabla \mu||d\vec \eta_\bot|,
\end{equation}
and, 
\begin{equation}
|d\vec \eta_\bot|={d\mu\over |\vec \nabla \mu|}.
\end{equation}
Substituting in Eq. \ref{set} and taking into account that ${\rm PDF}(\mu_0)d\mu\propto \sum_i dS_i(\mu_0)$, we finally, obtain,

\begin{equation}
\label{set2}
{\rm PDF}(\mu_0)\propto \sum_i \int_{\vec \eta_i(l,\mu_0)}{dl\over  |\vec \nabla \mu|}.
\end{equation}
{\bf We can use this expression to derive the PDF of a linear superposition of independent point masses (see \S \ref{discretization}).
In dimensionless units, $\vec y=\vec \eta /\eta_0$, where $\eta_0$ is the Einstein radius projected in the source plane,  the radial dependence of the total magnification of a point source induced by a single point mass is given by,

\begin{equation}
\label{mag0}
\mu(y^1,y^2)={y^2+2 \over y\sqrt{y^2+4}},
\end{equation}
where $y=\sqrt{(y^1)^2+(y^2)^2}$ is the radial distance to the microlens. The equation to calculate PDF($\mu$) (Eq. \ref{set}) written in dimensionless units is,

\begin{equation}
\label{less}
{\rm PDF}(\mu)={1\over A_y} \int_{\vec y(l,\mu)}{dl\over  |\vec \nabla \mu|},
\end{equation}
where $A_y$ is the magnification map area. In this case $\mu$ is discontinuous only at the origin and, consequently, we need only one open set to cover all the magnification map excluding the origin. The curves of constant $\mu$,  $\vec y(l,\mu)$ are circles of length, 
%
%
%
%
%Thus, the continuous magnification map is given by,
%
%\begin{equation}
%\mu(y^1,y^2)={y^2+2 \over y\sqrt{y^2+4}}.
%\end{equation}
%The probability density of a given value of $\mu$ is proportional to the infinitesimal surface element corresponding to a given value of $\mu$,

\begin{equation}
\label{a}
l(\mu)=2\pi y(\mu).
\end{equation}
Inverting Eq. \ref{mag0} for $\mu \ge 1$ we have,

\begin{equation}
\label{b}
y(\mu)= \sqrt{2} \sqrt{-1+\sqrt{\mu^2\over (\mu^2 - 1)}},
\end{equation}
and,

\begin{equation}
\label{c}
dy=-{1 \over (\mu^2 - 1)^{3\over 2} \sqrt{2} \sqrt{-1 +\sqrt{\mu^2\over (\mu^2 - 1)} }}d\mu.
\end{equation}
On the other hand, as  $|\vec \nabla \mu|$ is constant along each $\vec y(l,\mu)$ circle,
\begin{equation}
\label{less2}
{\rm PDF}(\mu)={1\over A_y} \int_{\vec y(l,\mu)}{dl\over  |\vec \nabla \mu|}={1\over A_y}{l\over |{d\mu \over dy}|}.
\end{equation}
Substituting from Eqs. \ref{a}, \ref{b} and \ref{c}, we obtain the PDF of a single point lens,

\begin{equation}
\label{less3}
{\rm PDF}(\mu)={1\over A_y} {2 \pi \over(\mu^2 - 1)^{3/2}}.
\end{equation}
%\begin{equation}
%dS(\mu)=-2 \pi {1\over(\mu^2 - 1)^{3/2}}d\mu. 
%\end{equation}
If we multiply by the total number of microlenses, ${\kappa \over \pi} A_x$, where $\kappa$ is the projected mass density in dimensionless units, and $A_x=\int\int dx^1dx^2$ is the area of the image plane mapped onto $A_y$, we can obtain the PDF for a linear superposition of point masses,

\begin{equation}
\label{final}
{\rm PDF}(\mu)=2 \langle \mu\rangle {\kappa\over(\mu^2 - 1)^{3/2}}, 
\end{equation}
making use of the definition of mean magnification, $\langle\mu\rangle$, as the ratio between mapped areas.}
}
\section{Magnification map computations\label{maps}}

When the mass ratio of the components is very large ($\sim$ 100), the computation of a magnification map corresponding to a bimodal distribution of microlenses  is not straightforward. In fact, computed magnification maps show a tendency to have a mean magnification smaller than the one theoretically expected. This aspect may not be relevant for many applications but can become very important when comparing {\it via} a $\chi^2$ test our mock and model PDFs.

To analyze this problem we make some simulations taking progressively larger circular regions for the stellar distribution and removing the borders of the maps before computing the mean magnification. We find that when there is a small number of microlenses, those that are close to the borders sometimes "pull out" rays (i.e. bits of the image plane area) that would have fallen into the magnification map if the distribution of matter were smooth, and sometimes "put" in the map rays that would have fallen outside of the limits of the map in the case of the smooth matter distribution.  As far as the size of the shooting region oversizes by a large factor the region of the image plane that is mapped onto the magnification map in the case of the smooth matter distribution, this would not be a problem: in  some maps the magnification would be greater and in other smaller than the theoretical one but the average of the histograms will converge to the theoretical one. However, if the oversize is not large enough\footnote{The needed oversize depends on the mass of the microlenses.} the rays diverted outside the map will not be compensated by the rays diverted into the map. This aspect is specially critical when you have a bimodal distribution with very small and very big microlenses, because you will tend to have a small number of big lenses and the fluctuations will be considerably larger. Thus the solution is to have large circular distributions of microlenses and large shooting regions. 

%We have checked that the mean magnification of the computed maps match the theoretical one with a deviation less than {\color{red} a few percent}.

%To fulfill these requirements we oversize by a factor {\color{red}  XXX} with respect to the map side the radius of the circular region where the microlenses are randomly distributed. 

For  the case ($\kappa=0.55$, $\gamma=0$),  we have computed, using the Inverse Polygon Mapping technique ({\bf IPM, }Mediavilla et al. 2006, 2011), maps of  {3.78 $\times$ 3.78} Einstein radii and {3380 $\times$ 3380} pixels with microlenses distributed in a circle of  {17.82} Einstein radii of radius. In the case of the mass ratio equal to 100, that means  {87} BHs and {8732} stars. The shooting region size extends  {12.6 $\times$ 12.6} Einstein radii. We remove  {200} border pixels from the maps to avoid edge effects. Each histogram is the result of the sum of  {200} histograms to mitigate the effects of sample variance, which could make all the study useless.

The situation is more complicated in the high magnification ($\kappa=0.55$, $\gamma=0.55$) case, for the shooting region size needs to be significantly larger. In this case we have also computed maps of  {3.78 $\times$ 3.78} Einstein radii and {3380 $\times$ 3380} pixels but now with microlenses distributed in a circle of  {80.19} Einstein radii of radius. In the case of the mass ratio equal to 100, that means  {1768} BHs and {176819} stars. The shooting region size extends  {56.7 $\times$ 56.7} Einstein radii. We remove  {400} border pixels from the maps to avoid edge effects. Each average histogram is the result of the sum of {45} histograms. Although this number is smaller than the 200 histogram averaged in the ($\kappa=0.55$, $\gamma=0$) case, notice that the size of the image plane region mapped back onto the magnification map (and, hence, the number of microlenses) is significantly larger in the high magnification case, even improving the statistical significance of the final average histograms.

%We have used the Inverse Polygon Mapping technique (Mediavilla et al. 2006, 2011) to compute the magnification maps, considering a number of rays per pixel in  absence of lensing of {\color{red}  0.1}.

{\bf The computation of this huge quantity of large size and high resolution maps has been possible thanks to the use of IPM (Mediavilla et al. 2006, 2011). The concepts and parameters of IPM are similar to those of the more familiar technique of Inverse Ray Shooting (IRS). In both methods a congruent lattice of points (defining cells which tessellate the image plane) is backwardly transported to the source plane using inverse lens mapping. In the case of IRS, the magnification of each source-plane pixel is made proportional to the number of points ("rays") that hit the pixel. In the case of IPM, an algorithm based in Green's theorem is used to exactly apportioning the area of each cell among the source-plane pixels covered by the transformed cell. In this way, the critical quantity in terms of accuracy and computation time, the number of rays (cells) per unlensed pixel, can be drastically reduced by two orders of magnitude or more. The main drawback of the current IPM codes is the lack of an hierarchical-tree algorithm (available in many IRS based routines) to avoid the linear dependence of computation time with the number of microlenses.}

\clearpage
\begin{figure}[h]
%\vskip -1 truecm
\includegraphics[scale=1.25]{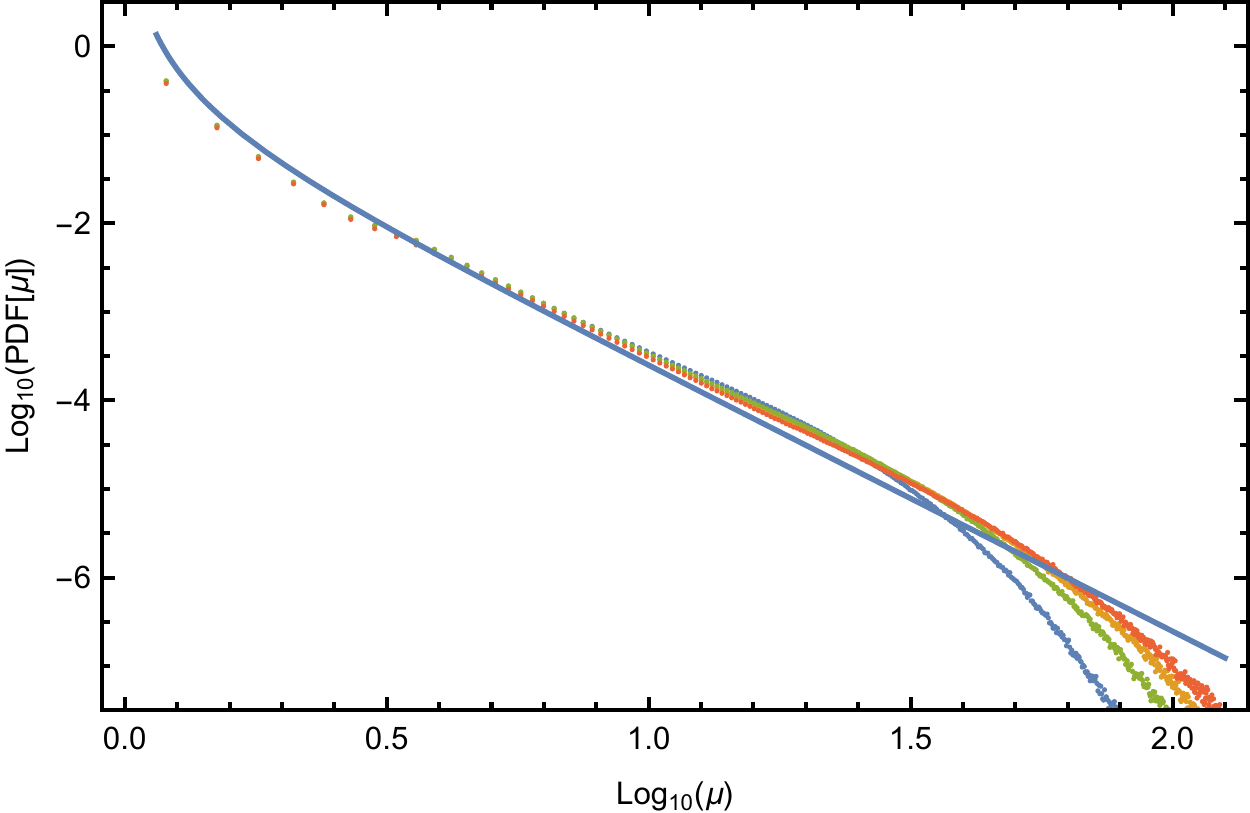}
%\plotone{chi2.pdf}
\caption{Data points are the numerical PDFs corresponding to $\kappa=0.1$ for four different masses with ratios $1:0.6:0.3:0.1$ (red : orange : green : blue). The blue continuous curve is the linear superposition model (see text). Notice the smoothness of the numerical PDFs, which are the average of 500 histograms to remove the effects of sample variance. \label{linear}}
\end{figure}

\clearpage
\begin{figure}[h]
%\vskip -1 truecm
\includegraphics[scale=0.09]{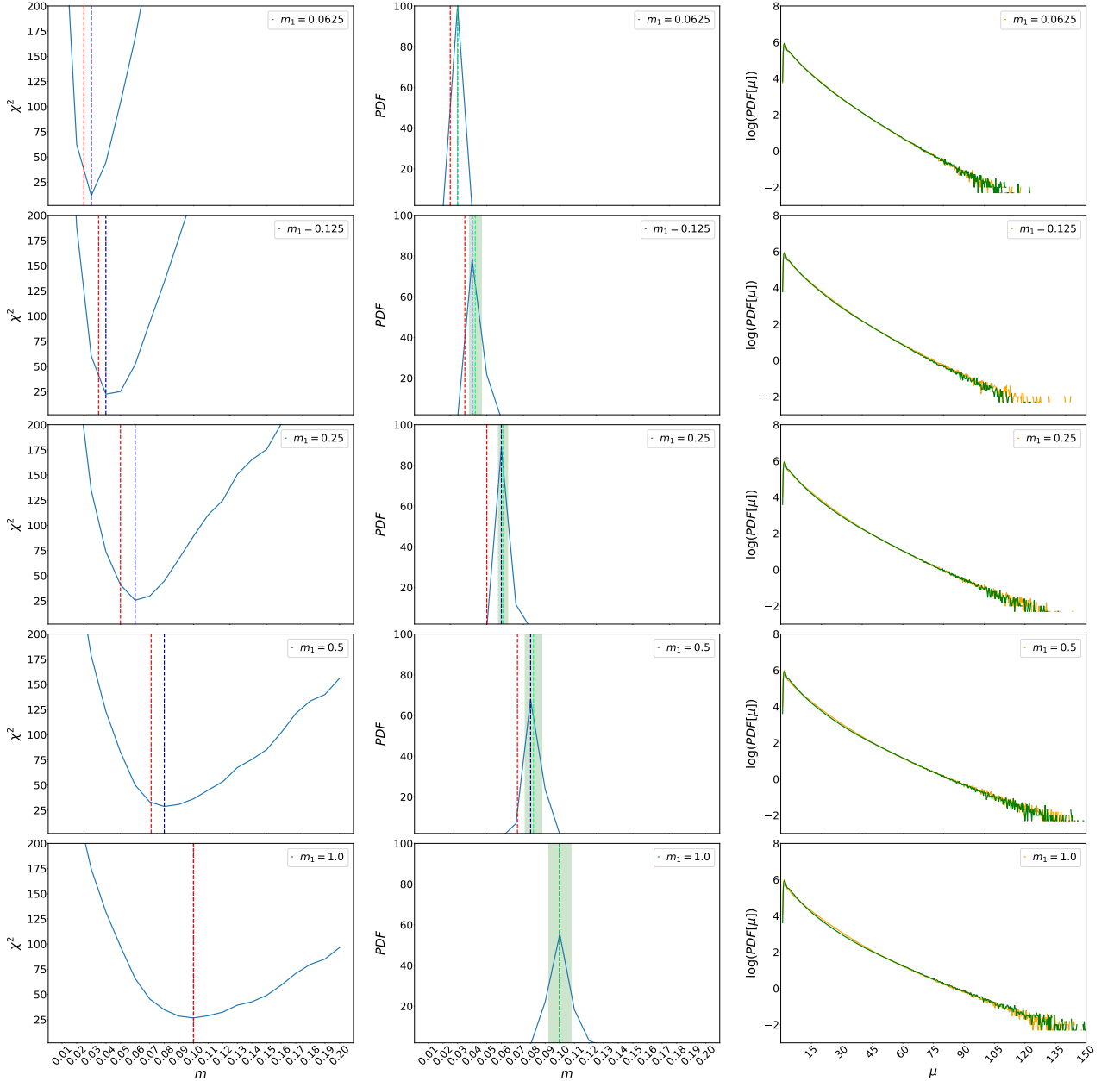}
%\plotone{chi2.pdf}
\caption{From top to bottom panels correspond to  a BH mass (in arbitrary units): $m_1=0.0625,0.125,0.25,0.5,1$. We adopt a mass for the stellar population, $m_2=0.01$, so that $m_1/m_2=6.25,12.5,25,50,100$. Left column: $\chi^2$, vertical lines mark the minimum (dark blue) and the GM (red). Middle column: PDFs inferred from $\chi^2$ with $\pm 1 \sigma$ intervals in green. Right column: mock (yellow) and model best-fit (green) histograms. (See text). \label{all}}
\end{figure}

%\begin{figure}[h]
%%\vskip -1 truecm
%\includegraphics[scale=0.15]{chi2.pdf}
%%\plotone{chi2.pdf}
%\caption{$\chi^2$. \label{chi2}}
%\end{figure}
%\clearpage
%\begin{figure}[h]
%%\vskip -1 truecm
%\includegraphics[scale=0.15]{bestfit.pdf}
%%\plotone{bestfit.pdf}
%\caption{Best fits.\label{bestfit}}
%\end{figure}
%\clearpage
%\begin{figure}[h]
%\includegraphics[scale=0.15]{PDF.pdf}
%%\plotone{PDF.pdf}
%\caption{PDFs. \label{PDF}}
%\end{figure}
%\clearpage
\begin{figure}[h]
\includegraphics[scale=0.5]{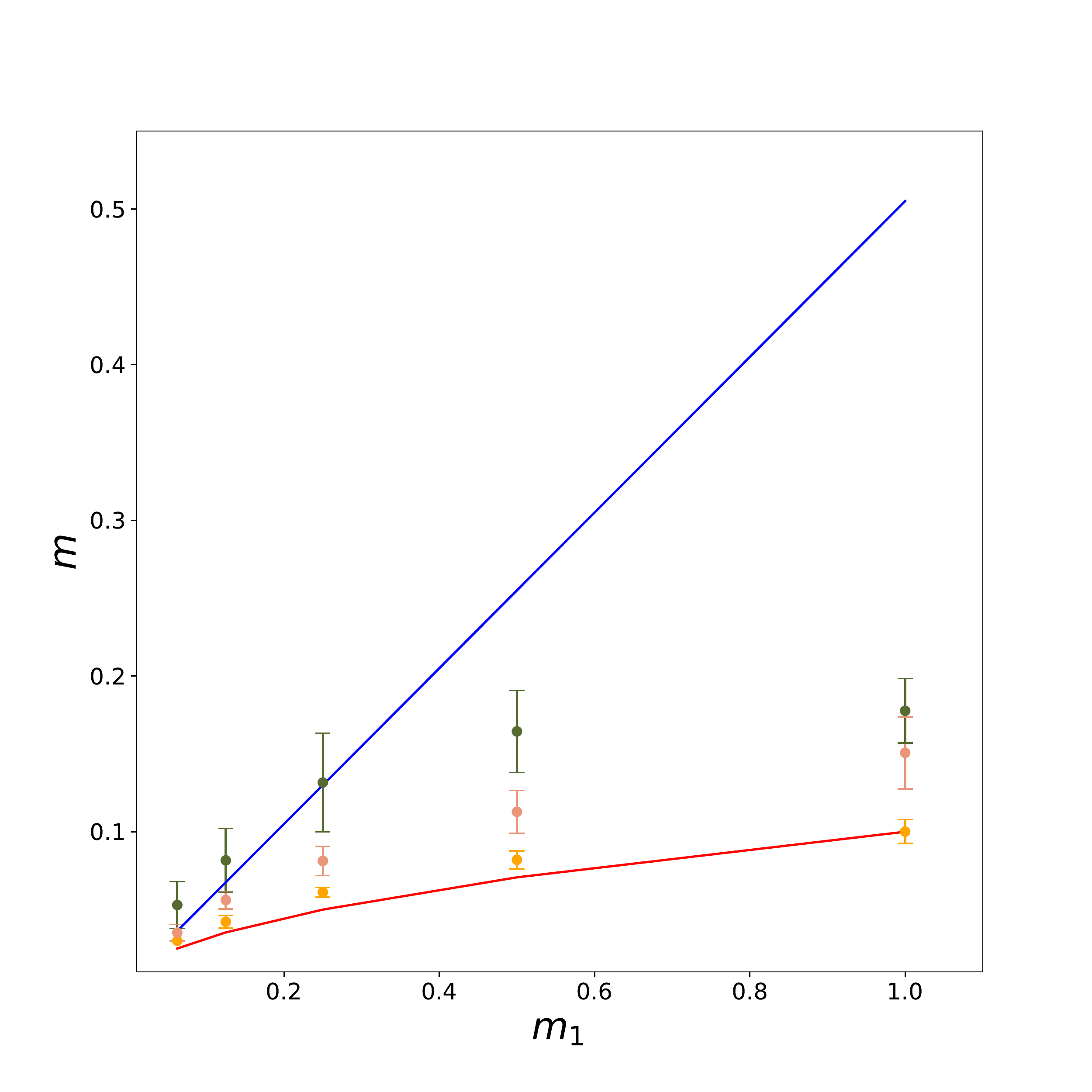}
%\plotone{mock_error.pdf}
\caption{Case $\kappa=0.55,\ \gamma=0$. We represent the best fit mass, $m$, versus the BH mass, $m_1$ in arbitrary units. (We adopt a mass for the stellar population,  $m_2=0.01$, so that $m_1/m_2=6.25,12.5,25,50,100$). Data points correspond to the best fit masses, $m$, for three different resolutions: 3380 pixels $\times$ 3380 pixels (yellow), 1690 pixels $\times$ 1690 pixels (orange) and 845 pixels $\times$ 845 pixels (green). The red (blue) curve is the GM (AM). See text.\label{mock_error}}
\end{figure}

\clearpage
\begin{figure}[h]
\includegraphics[scale=0.5]{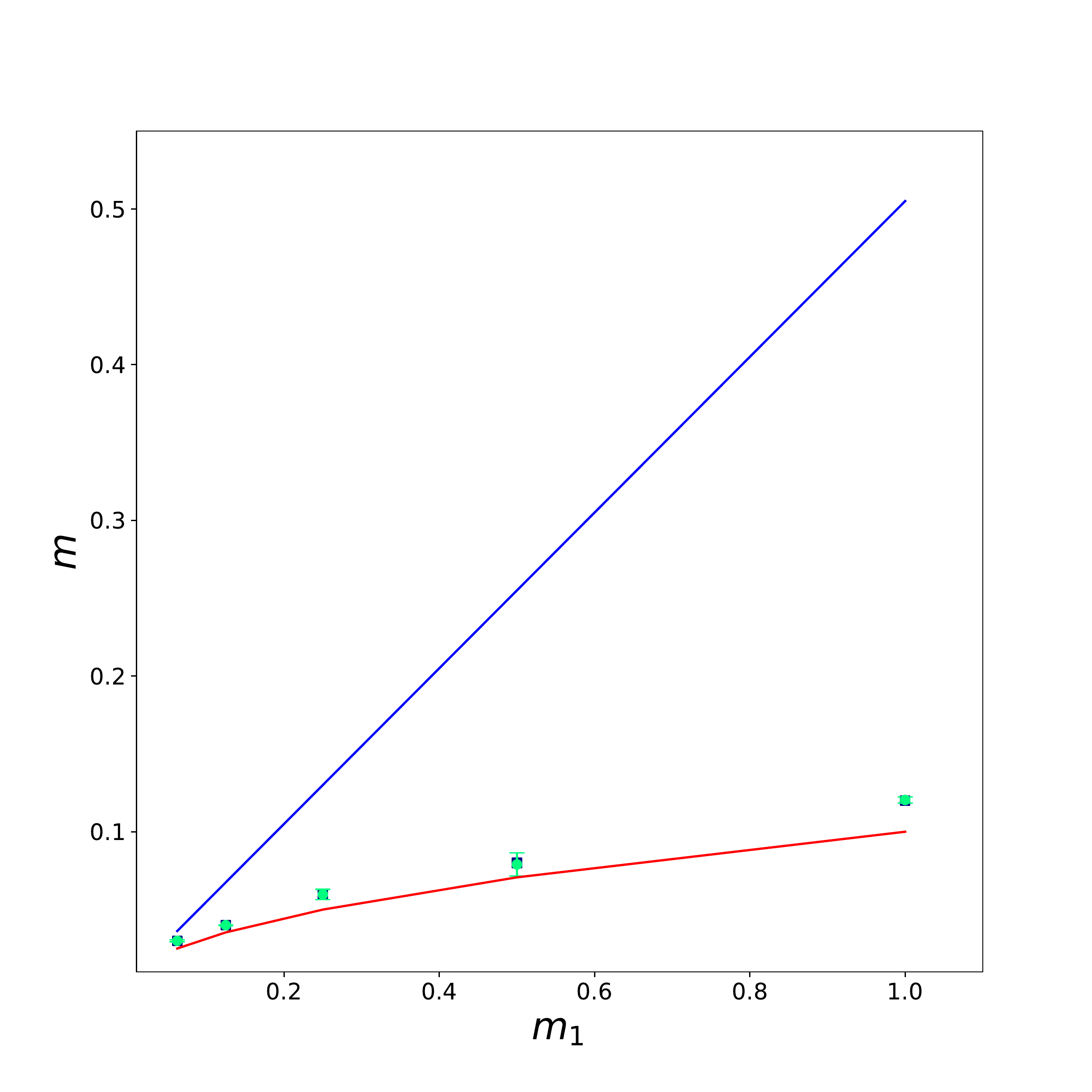}
%\plotone{mock_error.pdf}
\caption{As Figure \ref{mock_error} but for the case $\kappa=\gamma=0.55$. Data points correspond to the best fit masses, $m$, for a 3380 pixels $\times$ 3380 pixels resolution. The red (blue) curve is the GM (AM). See text. \label{mock_gamma}}
\end{figure}

\clearpage

%\begin{table}
%%\centering
%\caption{Fit Parameters for the three HME.}
%\medskip
%\begin{tabular}{cccc}
%\hline
%Event & $\Delta t (days)$ & $\langle \sigma_{adj} \rangle$ & $\chi^2_{red}$ \\
%\hline
%A / JD 1500& $51.5^{+3}_{-3}$ & 0.06 & 1.52 \\
%A / JD 4000 & $60.0^{+9}_{-8}$ & 0.02& 3.6 \\
%C / JD 1360& $58.0^{+5}_{-5}$ & 0.05 &1.82 \\
%\hline
%\end{tabular}
%\end{table}

\end{document}